\documentclass[twocolumn,superscriptaddress]{revtex4-2}
\usepackage[colorlinks=true,citecolor=blue,linkcolor=blue,urlcolor=blue,]{hyperref}
\usepackage{graphicx, nicefrac, textcomp}
\usepackage[normalem]{ulem}
\usepackage{changes}
\usepackage{chngcntr}
\usepackage{amssymb}
\usepackage{soul}
\usepackage{color}
\usepackage{float}
\usepackage{textcomp}
\usepackage{amstext}
\usepackage{graphicx}
\usepackage{gensymb}
\usepackage{graphics}\usepackage{subfigure}\usepackage{longtable}\usepackage{pstricks}\usepackage{dcolumn}\usepackage{bm}
\usepackage{siunitx}

\begin{document}

\title{Realization of a classical Ruddlesden Popper type bilayer nickelate in Sr$_3$Ni$_{2-x}$Al$_{x}$O$_{7-\delta}$ with unusual Ni$^{4+}$}

\author{H.~Yilmaz}
\email[]{hasan.yilmaz@imw.uni-stuttgart.de}
\affiliation{University of Stuttgart, Institute for Materials Science, Materials Synthesis Group, Heisenbergstraße 3, 70569 Germany}
\author{K.~K\"uster}
\affiliation{Max Planck Institute for Solid State Research, Heisenbergstra{\ss}e 1, 70569 Stuttgart, Germany}
\author{U.~Starke}
\affiliation{Max Planck Institute for Solid State Research, Heisenbergstra{\ss}e 1, 70569 Stuttgart, Germany}
\author{O.~Clemens}
\affiliation{University of Stuttgart, Institute for Materials Science, Materials Synthesis Group, Heisenbergstraße 3, 70569 Germany}
\author{M.~Isobe}
\affiliation{Max Planck Institute for Solid State Research, Heisenbergstra{\ss}e 1, 70569 Stuttgart, Germany}
\author{P.~Puphal}
\email[]{puphal@fkf.mpg.de}
\affiliation{Max Planck Institute for Solid State Research, Heisenbergstra{\ss}e 1, 70569 Stuttgart, Germany}

\date{\today}

\begin{abstract}
The discovery of 80 K superconductivity in bilayer La$_3$Ni$_2$O$_7$ at pressures greater than 14 GPa presents a unique opportunity to study a novel class of high-temperature superconductors. Therefore, other bilayer nickelate following the classical ($T^{4+}$) Ruddlesden-Popper (RP) series of Sr$_3$Ni$_2$O$_7$ would present an interesting new candidate. In this work, we study the stabilization of RP $n=2$ phase in Sr$_3$Ni$_{2-x}$Al$_{x}$O$_7$, via floating zone growth of crystals. With powder and single crystal XRD, we study the stability range of the RP-type phase. Our Thermogravimetric Analysis (TGA), X-ray photoelectron spectroscopy (XPS) and gas extraction studies reveal a remarkably high oxidation state of Ni$^{4+}$. The obtained black crystals are insulating in transport and show a magnetic transition around 12 K.
\end{abstract} 

\maketitle

\section*{Introduction}
\begin{figure}[tb]
\centering
\includegraphics[width=1.0\columnwidth]{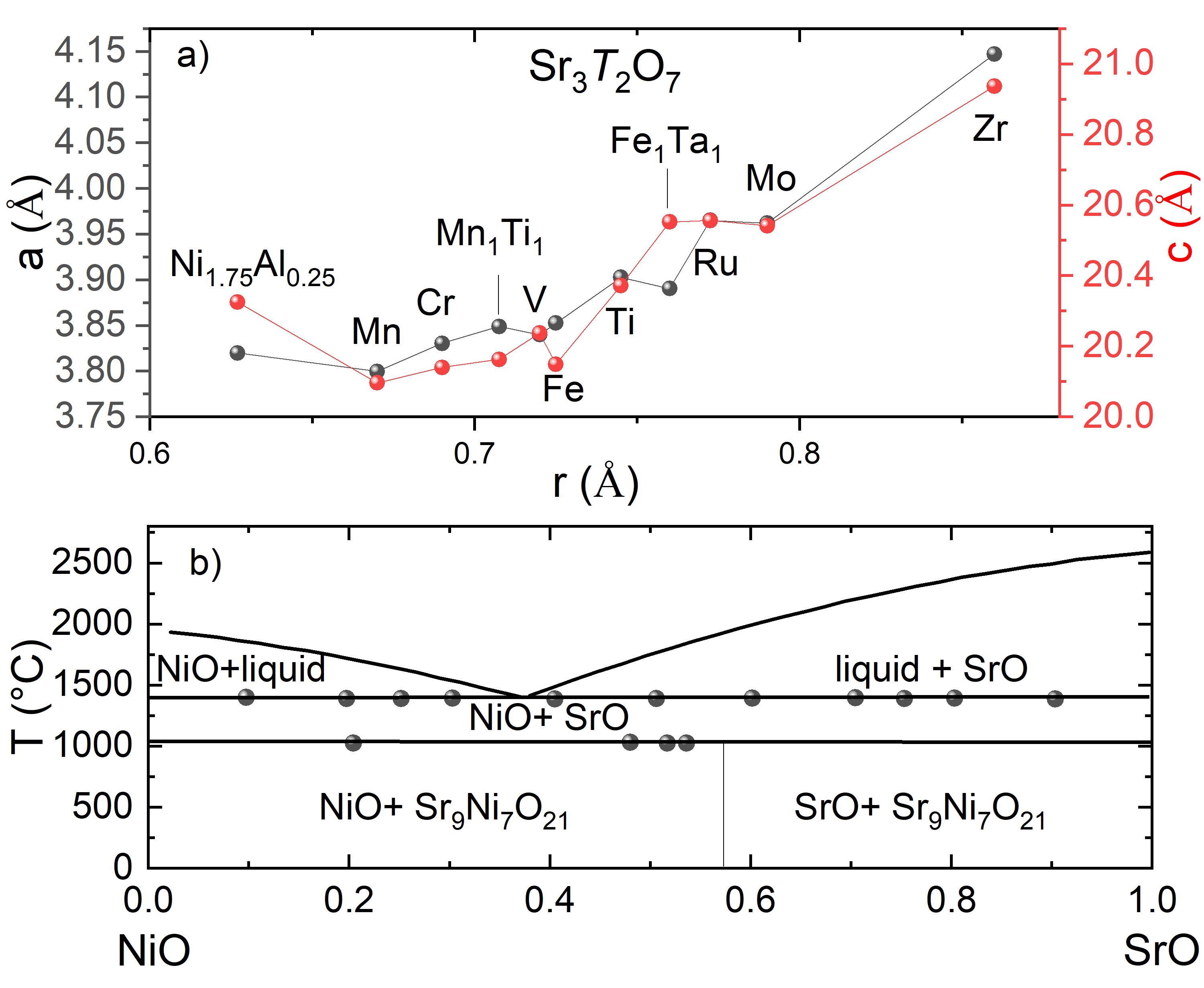}
\caption{\textbf{Literature status on Sr based Ruddlesden Popper nickelates.} a) Lattice constants (a,c) versus ionic radius (r) of the existing bilayer RP-type Sr$_3T_2$O$_7$ phase with $T=$ transition metal from Ref. \cite{Mitchell1998,CastilloMartinez2007,Sharma2004,Itoh1991,Dann1992,Elcombe1991,Mueller‐Buschbaum1990,Li2006,Li2020,Fukasawa2023}, with the nickelate phase of Ref. \cite{Kharlamova2018}. b) Phase diagram of Sr-Ni-O (in air $\rho_{O2}=0.209$ atm) digitized from Ref. \cite{Zinkevich2004}.
}
\label{RP}
\end{figure}

The Ruddlesden-Popper (RP) series denotes structural building blocks in oxide systems consisting of alternating rocksalt and perovskite layers with the composition $A_{n+1}T_n$O$_{3n+1}$, where $n$ is the number of consecutive perovskite blocks of $T$O$_6$ octahedral layers within a structural unit. 
In its most common form the $A$ site is found to be Strontium, while $T$ are various transition metal ions both 3d and 4d (see Fig. \ref{RP} extracted from Ref. \cite{Mitchell1998,CastilloMartinez2007,Sharma2004,Itoh1991,Dann1992,Elcombe1991,Mueller‐Buschbaum1990,Li2006,Li2020,Fukasawa2023}). With this $A$ = Sr the series denotes a clear oxidation state of $T$ with 4+ for stoichiometric oxygen content O$_{3n+1}$. Due to the flexibility of the oxidation state of transition metals $T$ a reduction down to Sr$_3T_2$O$_5$ is possible for certain transition metals $T$. Similarly, the $A$ site can be substituted by $A=RE$ a rare earth ion, leading to a lower oxidation state of $T$. However, only for Ni a full $A=$ La phase is reported for the whole RP phase series, which leads to a series of different oxidation states varying with n.

Nickelates have gained attention since 2019 with the discovery of superconductivity in infinite-layer nickelates \cite{Li2019,Nomura2022,Hepting2021,Chen2022,Held2022}, where Ni ions exhibit a formal 3$d^{9}$ configuration, isoelectronic and isostructural compared to cuprate superconductors \cite{Keimer2015,Lee2006}. 
Even though investigations of the $RE_{n+1}$Ni$_n$O$_{3n+1}$ compounds ($RE$ = rare-earth) started long ago \cite{Greenblatt1997,RodriguezCarvajal1991,Torrance1992}, recent high-pressure studies reported superconductivity both in La$_3$Ni$_2$O$_7$ \cite{Sun2023,Puphal2024} and La$_4$Ni$_3$O$_{10}$ \cite{sakakibara2023,Li2023La4310,Zhang2023La4310,Zhu2023La4310} with a $T_c$ of around 80 and 20 K, respectively. Unlike classical RP with $T^{4+}$
Ni in La$_3$Ni$_2$O$_7$ possesses an average 3$d^{7.5}$ configuration and 3$d^{7.33}$ in La$_4$Ni$_3$O$_{10}$. This is rather close to the formal 3$d^{7}$ state in perovskite nickelates, which were proposed to exhibit a cuprate-like electronic structure when sandwiched between layers of an insulator \cite{Chaloupka2008,Hansmann2009,Boris2011}. Furthermore, an average 3$d^{7.5}$ state is reminiscent of hole-doped (La,Sr)$_2$NiO$_4$, where 3$d^{7}$ and 3$d^{8}$ sites coexist \cite{Uchida2012,Uchida2011PRB,Uchida2011PRL}.

\begin{figure*}[tb]
\centering
\includegraphics[width=2.0\columnwidth]{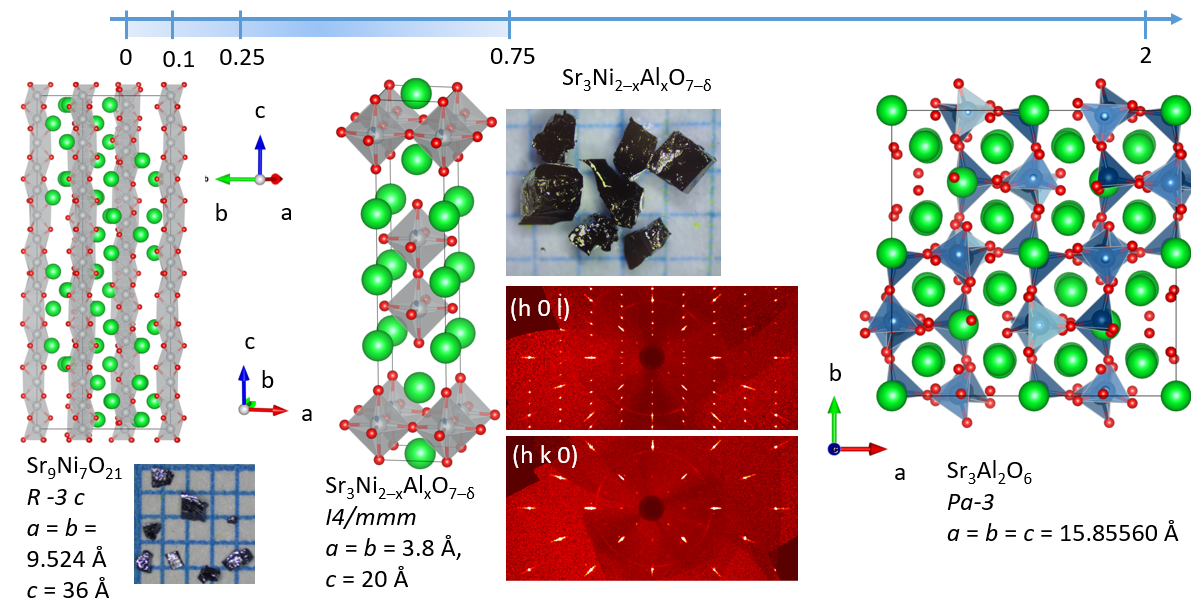}
\caption{\textbf{Doping dependence of the series Sr$_3$Ni$_{2-x}$Al$_{x}$O$_7$}, following the exploration of phase formations from Optical Floating Zone (OFZ) growth at 9 bar oxygen partial pressure shown in Fig. \ref{XRD}. The realized structures are shown with Ni (gray), Al (blue), Sr (green), and O (red). Two pictures of crystals extracted from the boule after breaking the phases Sr$_9$Ni$_{7}$O$_{21}$ (left) and Sr$_3$Ni$_{2-x}$Al$_{x}$O$_7$ (right) are shown with a millimeter grid paper underneath.  In the center below the crystal images, reconstructed maps of XRD intensities of the Sr$_3$Ni$_{1.75}$Al$_{0.25}$O$_7$ single crystals are shown.
}
\label{phases}
\end{figure*}

Theoretical investigations indicate that a strong intrabilayer coupling is a key ingredient in the superconducting mechanism, but the approaches to disentangle the intermixed multiorbital states in La$_3$Ni$_2$O$_7$ vary. Specifically, the roles assigned to low-energy Ni $3d_{x^2-y^2}$, $3d_{z^2}$, and O 2$p$ orbital degrees of freedom diverge among the models, yielding singlet-like states with partial analogies to those of 3$d^{9}$ cuprates \cite{Sun2023,Qin2023,Yang2023singlet,Wu2023ZRS} or different notions \cite{Luo2023,Christiansson2023,Lechermann2023,Nakata2017,Gu2023,chen2023,Shilenko2023,Lu2023Interplay,Zhang2023dimer,Oh2023,Lange2023}. 
Notably it was found that there is polymorphism in La$_3$Ni$_2$O$_7$ with a novel hybrid RP structure alternating of n = 1 and n = 3  \cite{Puphal2024}. The appearance of polymorphism limits the quality and size of the crystals in La$_3$Ni$_2$O$_7$ due to intergrowth of the two phases with a very narrow and close phase formation area \cite{Puphal2024}. 
This calls for additional RP-type nickelates to be studied in detail.

The first report on RP-type n = 2 Sr-based nickelate is published in Ref. \cite{Kharlamova2018} based on powder samples. They found that the substitution of Al$^{3+}$ into the Sr-Ni-O system stabilizes oxygen-deficient RP-type structures, in a stability range of 0.5 $\leq$ x $\leq$ 0.75.

Ni$^{4+}$ is a rare oxidation state, usually found with non-distorted octahedra, as shown for BaNiO$_3$ with a Ni-O distance of $d=1.8650(13)$ \AA, 
Li$_{2-x}$NiO$_3$ $d=1.88$ \AA \, \cite{Bianchini2020}, or Li$_x$(Ni,Co,Mn)O$_3$ $d=1.9$ \AA \, \cite{Ohnuma2019}. Notably, both Li related compound showed that via the control of Li$_x$, changing the oxidization of Ni, the bond distance varied from  $\approx2.1$ \AA \ for Ni$^{2+}$, to 1.9 \AA \, for Ni$^{4+}$.

In this article, we examine Sr$_3$Ni$_2$O$_7$ single crystal formation via Al substitution, establishing a new candidate of RP-type nickelates. We confirm that the parent phase without Al does not exist as an RP-type compound even working in enhanced oxygen pressure. We find an optimal synthesized compound for $x=0.25$ in Sr$_3$Ni$_{2-x}$Al$_{x}$O$_7$ in stark contrast to Ref. \cite{Kharlamova2018}, where a minimum of $x=0.5$ is found. Our powder and single crystal x-ray diffraction (XRD) experiments combined with x-ray photoemission spectroscopy (XPS), thermogravimetric analysis (TGA) and gas extraction confirm an extremely rare high oxidation state of Ni$^{4+}$. Most importantly, these crystals form homogenously, neither containing impurity phases (within the resolution limits) nor showing polymorphism. 

\section*{Results}
\subsection*{Synthesis}
As shown in Fig. \ref{RP} b) the binary phase diagram of SrO-NiO taken from Ref. \cite{Zinkevich2004} reveals besides SrO, Sr$_9$Ni$_7$O$_{21}$, and NiO no RP-type phase. In accordance, for our $x=0$ synthesis we find no stability of a RP-type phase, but instead obtain Sr$_9$Ni$_7$O$_{21}$ single crystals. In Tab. \ref{tab:SNO_OFZ} structural details are summarized. However with the introduction of Aluminium on the Ni site, a RP $n=2$ phase emerges. In Fig. \ref{phases} we display the phases formed for varied $x$, where the top panel shows the $x$ content, and below are the crystal structures shown as well as single crystal pieces. Most interestingly even the pure SrO-Al$_2$O$_3$ forms no RP-type phases. However, for $x\geq0.1$, a mix of the RP-type phase and Sr$_9$Ni$_7$O$_{21}$ occurs.

\subsection*{Phases and structure}
\begin{figure}[tb]
\centering
\includegraphics[width=1.0\columnwidth]{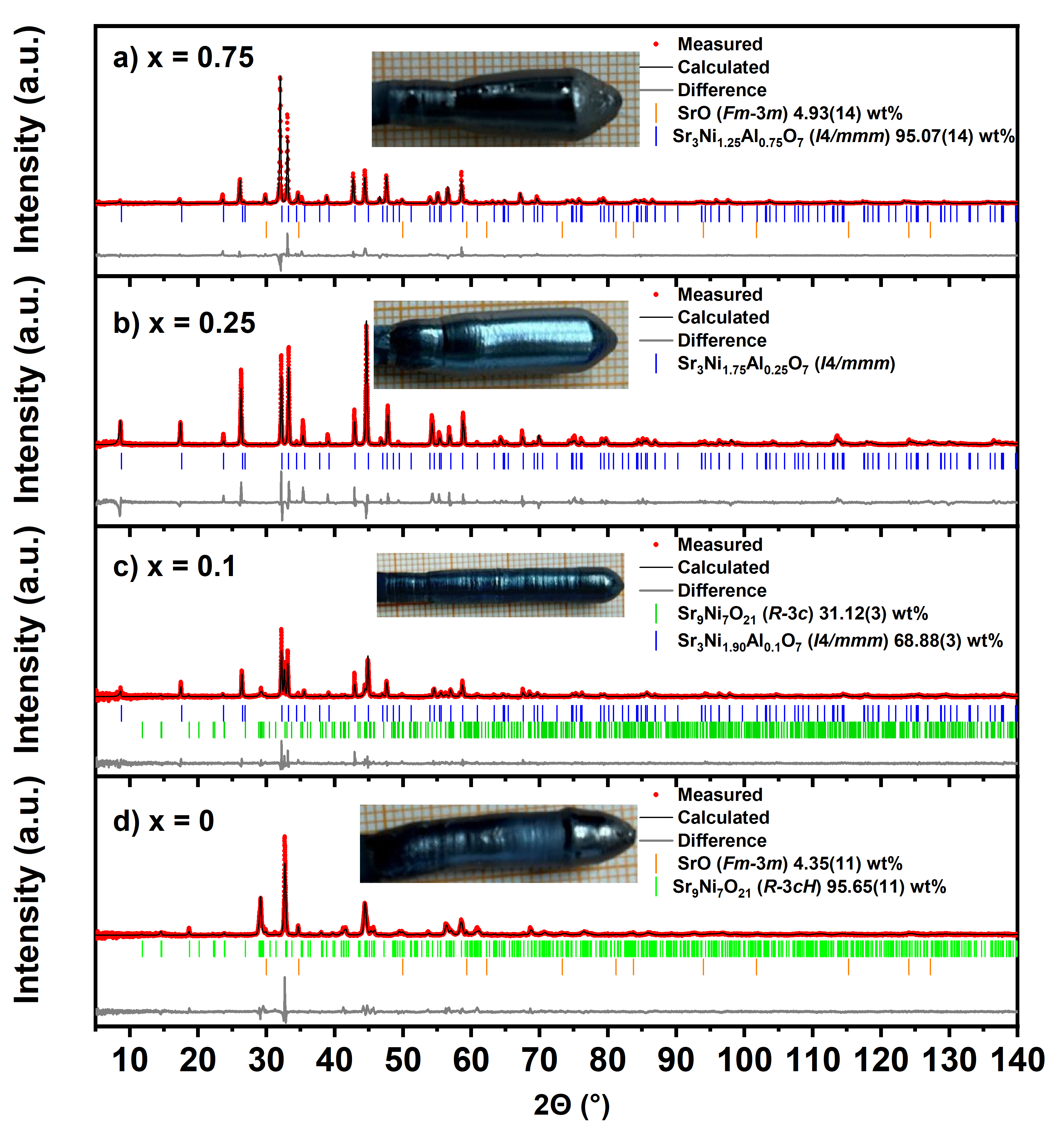}
\caption{\textbf{Rietveld refinement of the room-temperature Powder X-ray diffraction pattern of Sr$_3$Ni$_{2-x}$Al$_{x}$O$_7$} for a) x=0.75 ; b) x=0.25 c) x=0.10 d) x=0. The solid black line corresponds to the calculated intensity from the Rietveld refinement, the solid gray line is the difference between the experimental and calculated intensities, and the vertical colored bars are the calculated Bragg peak positions. The underlying phases are given in the legend for phase identification of four distinct floating zone growth with pictures of the corresponding boules as an inset, with a millimeter scale underneath.
}
\label{XRD}
\end{figure}

The corresponding XRD powder patterns of Sr$_3$Ni$_{2-x}$Al$_{x}$O$_7$ samples are shown in Fig. \ref{XRD}. As can be clearly seen from Fig. \ref{XRD} a), b), and c), the compounds show a high phase purity containing only (mainly) a $n=2$ RP-type compound with $I4/mmm$ (\#139) symmetry. In accordance with the general RP-type n=2 phase \cite{beznosikov2000perovskite}, Sr atoms (green) are placed in one of the large 12-coordinate sites in perovskite layers to oxygen (red) or one of the smaller 9-coordinate sites in rocksalt layers. As depicted in Fig. \ref{phases} Ni/Al (grey/blue) atoms are positioned inside anionic octahedra. The Ni-O distance of the outer oxygen amount to
$d=1.89(3)$ \AA \, while the inner oxygen extend to $d=1.984(5)$ \AA  and the in-plane distance lies close to outer with 1.9038(9) \AA. As discussed in the Introduction, compared to Li-based nickelates this shows that except for the elongated inner bond, it falls well into the expected distance for Ni$^{4+}$ with 1.9 \AA \, \cite{Bianchini2020,Ohnuma2019}. This is quite different from the related RP La$_3$Ni$_2$O$_7$, which seems to show enormous distortions with the outer apical oxygen having a distance of 2.3 \AA \, \cite{Sun2023}.

\begin{table}[h]
\caption{Refined atomic coordinates of a float zone grown boule of the stoichiometry x=0: "Sr$_{3}$Ni$_{2}$O$_{7}$" found to crystallize in the trigonal $R$-3$c$ (\#167) structure of Sr$_{9}$Ni$_{7}$O$_{21}$, extracted from powder XRD ($a=9.4527(4)$ {\AA}, $c=36.004(2)$ {\AA}, Rwp$=6.45$, GOF$=1.42$)}
\label{tab:SNO_OFZ}
\begin{tabular}{cccccc}
\hline
\textbf{Atom} & \textbf{Site} & \textbf{x} & \textbf{y} & \textbf{z} & \textbf{Occ} \\ \hline
Sr & Sr01 & 0.3300(6) & 0.3388(6) & 0.69569(15) & 1\tabularnewline
Sr & Sr02 & 0.6766(4) & 0.6766(4) & 0.74431(17) & 1\tabularnewline
Ni & Ni1 & 1/3 & 2/3 & 2/3 & 1\tabularnewline
Ni & Ni2 & 2/3 & 1/3 & 0.6154(6) & 1\tabularnewline
Ni & Ni3 & 1/3 & 2/3 & 0.7387(6) & 1\tabularnewline
Ni & Ni4 & 1/3 & 2/3 & 0.6869(6) & 1\tabularnewline
O & O1 & 0.517(2) & 0.184(2) & 0.6303(8) & 1\tabularnewline
O & O2 & 0.490(4) & 0.675(3) & 0.6951(10) & 1\tabularnewline
O & O3 & 0.499(4) & 0.311(3) & 0.6464(9) & 1\tabularnewline
O & O4 & 0.141(3) & 0.667(4) & 0.7549(9) & 1\tabularnewline
 \hline
\end{tabular}
\end{table}

\begin{table*}[tb]
\caption{Refined atomic coordinates of Sr$_3$Ni$_{2-x}$Al$_{0x}$O$_7$ in the tetragonal $I4/mmm$ (\#139) structure, extracted from powder XRD refinements shown in Fig. \ref{XRD}. Given are the individual z sites and occupancy (occ) for the corresponding x sorted from 0.75 to 0.25 to 0.1. The corresponding lattice constants and quality parameters are: \newline\textbf{x=0.75}: $a=3.8126(6)$ {\AA}, $c=20.3393(4)$ {\AA}, Rwp$=5.08$, GOF$=2.59$; \newline \textbf{x=0.25}: $a=3.8020(5)$ {\AA}, $c=20.2616(3)$ {\AA}, Rwp$=8.80$, GOF$=2.25$; \newline  \textbf{x=0.10}: $a=3.8104(6)$ {\AA}, $c=20.1495(3)$ {\AA}, Rwp$=5.82$, GOF$=0.97$}
\label{tab:SNOatom_pos}
\begin{tabular}{cccccccccc}
\hline 
 &  &  &  & \multicolumn{2}{c}{``x = 0.75''} & \multicolumn{2}{c}{``x = 0.25''} & \multicolumn{2}{c}{``x = 0.1''}\tabularnewline
\textbf{Atom}  & \textbf{Site}  & \textbf{x}  & \textbf{y}  & \textbf{z}  & \textbf{Occ}  & \textbf{z}  & \textbf{Occ}  & \textbf{z}  & \textbf{Occ} \tabularnewline
\hline 
Sr  & Sr01  & 1/2  & 1/2  & 0.8165(6)  & 1  & 0.8170(4)  & 1  & 0.8174(6)  & 1 \tabularnewline
Sr  & Sr02  & 0  & 0  & 1/2  & 1  & 1/2 & 1 & 1/2 & 1 \tabularnewline
Ni  & Ni01  & 1/2  & 1/2  & 0.5988(13)  & 5/8  & 0.6012(11)  & 7/8  & 0.6009(14)  & 19/20 \tabularnewline
Al  & Al01  & 1/2  & 1/2  & 0.5988(13)  & 3/8  & 0.6012(11)  & 1/8  & 0.6009(14)  & 1/20 \tabularnewline
O  & O01  & 1/2  & 1/2  & 0.6937(4)  & 1  & 0.6934(3)  & 1  & 0.6899(3)  & 1 \tabularnewline
O  & O02  & 0  & 1/2  & 0.5922(5)  & 0.6333(8)  & 0.5911(16)  & 1  & 0.5882(3)  & 0.9130(7) \tabularnewline
O  & O03  & 1/2  & 1/2  & 1/2  & 1  & 1/2  & 1  & 1/2  & 1 \tabularnewline
\hline 
\end{tabular}
\end{table*}

 As mentioned above, we also investigated the undoped (x=0) sample and Fig. \ref{XRD} d) shows that it has been crystallizing Sr$_9$Ni$_7$O$_{21}$ in the trigonal $R$-3$c$ structure (see Tab. \ref{tab:SNO_OFZ}) with 4.35\% SrO impurities. As a result of this finding, we conclude that it is not possible to sythesize stoichiometric Sr$_3$Ni$_{2}$O$_7$ since the oxidation state +4 is not a common state for Ni and in perovskites even Ni$^{3+}$ require enormous oxidizing conditions \cite{Puphal2023APL} and already for PrNiO$_3$ with Ni$^{3+}$ 300 bar are required.
 The x = 0.10 boule contains two phases which are 68.88\% Sr$_3$Ni$_{2-x}$Al$_{x}$O$_7$ tetragonal $I4/mmm$ (results see Tab. \ref{tab:SNOatom_pos}) and 31.12\% Sr$_9$Ni$_7$O$_{21}$ trigonal $R$-3$c$ (depicted in Fig. \ref{phases} on the left).
In contrast, the XRD analysis of a crushed crystal of x=0.25 indicated a phase pure composition and also matched well with reports on powder samples with higher x \cite{Kharlamova2018}. The crystal structure was solved from the single-crystal XRD data, and then the determined structural model was used as a starting point for powder diffraction. The refined results are summarized in Tab. \ref{tab:SNOatom_pos}. Unlike the powder data from literature, which suggested a stability range from 0.5 to 0.75, the x = 0.75 crystal contained 4.93wt\% of SrO as an impurity phase with the refinement result summarized in Tab. \ref{tab:SNOatom_pos}, indicating a possible upper limit. The lower limit and upper limit are further discussed in the Chemical composition section.
Hence, we have discovered that stabilization of the Sr$_3$Ni$_{2}$O$_7$ Ruddlesden-Popper phase can be achieved by a small amount of Al$_2$O$_3$ substitution. In Tab. \ref{tab:SNOatom_pos} we list the lattice parameter versus the substitutional content which show a clear deviation for the $a$-axis from Vegards law below 0.25 in accordance with a separate phase evolving. We hence conclude a lower solubility limit close to 0.25.

\newpage

\subsection*{Chemical composition}

\begin{figure}[tb]
\includegraphics[width=1.0\columnwidth]{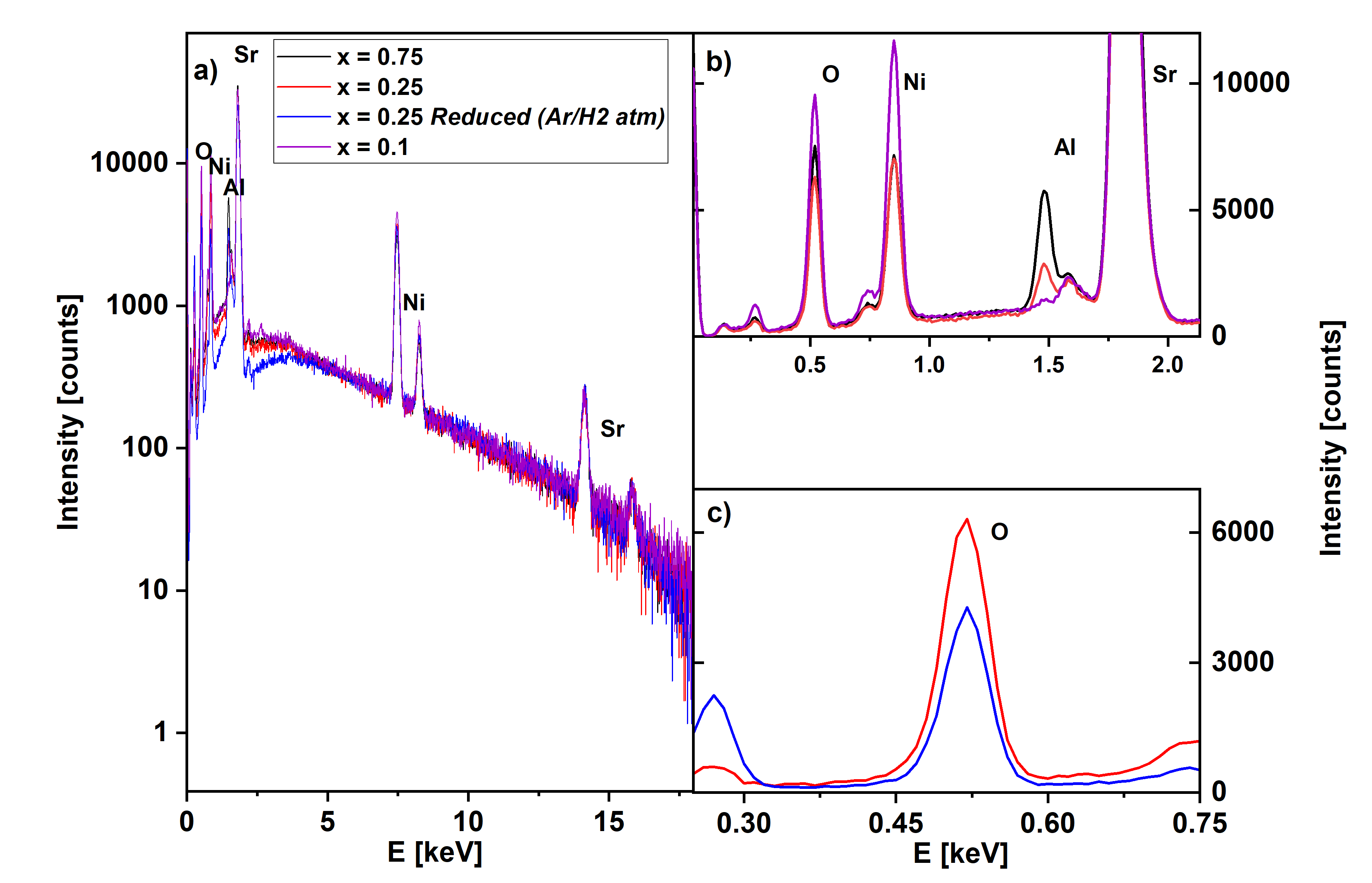}
\caption{\textbf{Stoichiometry:} a) Comparison of Energy-Dispersive X-ray (EDX) spectra in a logarithmic scale normalized to the Sr k$_\beta$ line for x=0.75, 0.25, 0.25 reduced, and 0.1 samples of the Sr$_3$Ni$_{2-x}$Al$_{x}$O$_7$ phases. b) Magnification of the low energy part visualizing the difference in Al doping for different x (notably the given spectra of x = 0.1 hints at a Sr$_9$Ni$_{7}$O$_{21}$ phase). c) Magnification of the oxygen content of x=0.25 for as grown (red) and reduced (blue).
}
\label{EDX}
\end{figure}

We have analyzed the stoichiometry of the single crystals with a combination of energy dispersive x-ray (EDX), thermogravimetric analysis (TGA) and gas extraction. Representative EDX spectra are plotted in Fig. \ref{EDX} for all compositions, as well as the reduced crystals. For the $x=0.75$ batch the average EDX results yield a slightly reduced Al content Sr$_{3.02(4)}$Ni$_{1.38(2)}$Al$_{0.57(1)}$O$_{5.5(1)}$, while oxygen from EDX is likely underestimated.
For the $x=0.1$ we find as expected from the lattice parameter plot that a solubility limit is reached. Individual EDX spectra show a reduced Al content (as the one given in Fig. \ref{EDX})  compared to the previous batch, but this is from local points of the impurity phase Sr$_9$Ni$_{7}$O$_{21}$. This is best captured in the fact that the obtained crystals show an average stoichiometry of Sr$_{2.99(7)}$Ni$_{1.75(6)}$Al$_{0.26(3)}$O$_{7(1)}$, which is an equally high Al content as $x=0.25$.

Finally, we focus on the optimal synthesis with $x=0.25$. Here, the Al content fits very well to the expected value Sr$_{3.03(1)}$Ni$_{1.81(2)}$Al$_{0.25(1)}$O$_{6.88(5)}$. Note that only for this batch the oxygen content was determined by gas extraction as described in the methods section. We highlight the experimentally determined exact balancing of the oxygen content stabilizing Ni$^{4+}$, with 12.5\% Al$^{3+}$. The annealing of this sample at 600 bar oxygen pressure did not enhance the oxygen content and none of the physical properties changed (as visible in our TGA results discussed in the following). Thus we conclude that this is the fully oxidized state. 

\begin{figure}[tb]
\includegraphics[width=1.0\columnwidth]{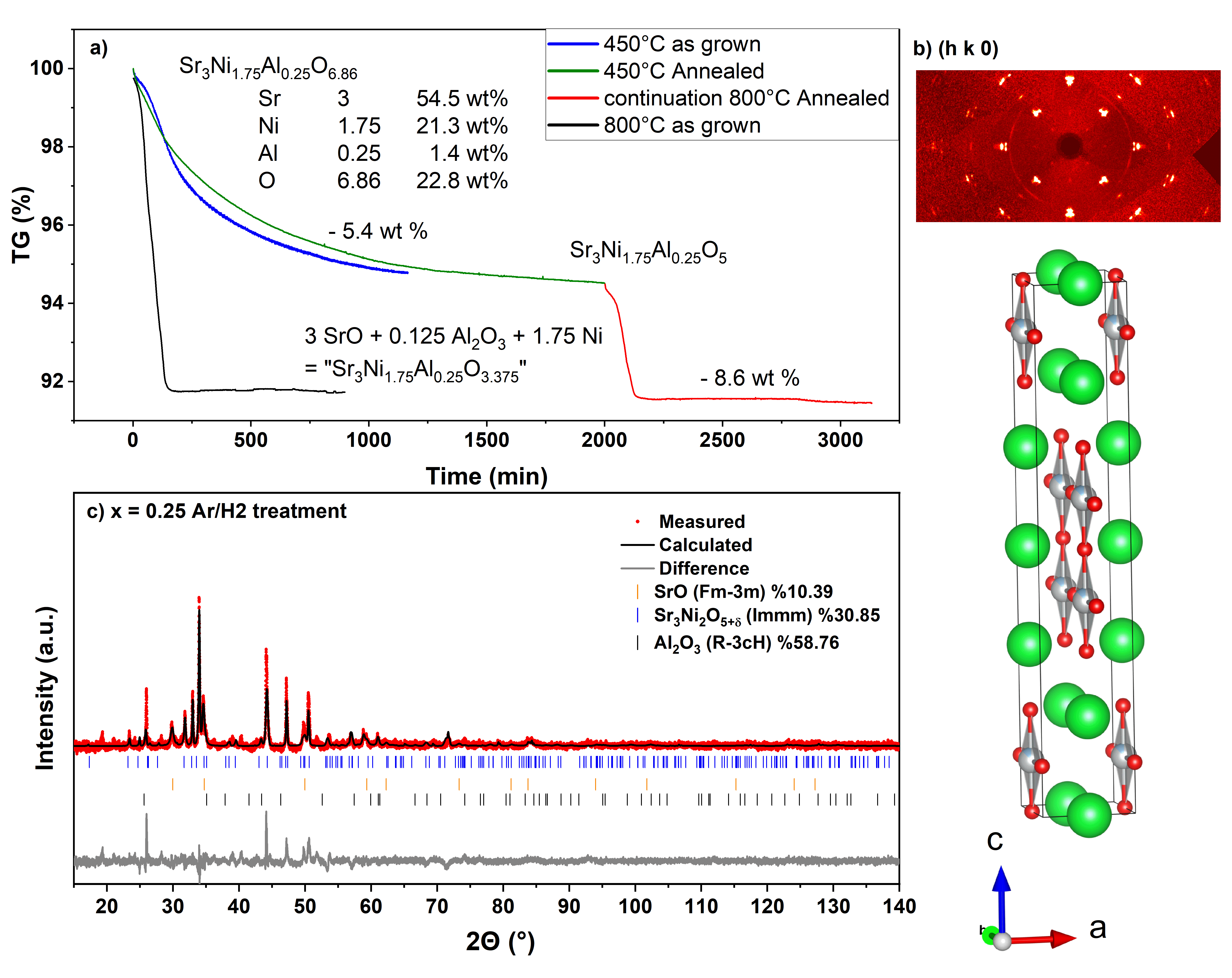}
\caption{\textbf{Topochemical oxygen reduction realized via} a) Thermogravimetric analysis (TGA) of the Sr$_3$Ni$_{1.75}$Al$_{0.25}$O$_7$ (x=0.25) single crystals both annealed and not annealed heated at 450$\degree$C and 800$\degree$C in Ar/5\%H$_2$ flow plotted versus time. b) Crystal structure derived from  XRD are shown with Ni(grey), Al (blue), Sr (green) and O (red). c) Rietveld refinement of the room-temperature XRD pattern of Sr$_3$Ni$_{1.75}$Al$_{0.25}$O$_7$ crushed single crystals after Ar/H$_2$ treatment. The solid black line corresponds to the calculated intensity from the Rietveld refinement, the solid gray line is the difference between the experimental and calculated intensities, and the vertical colored bars are the calculated Bragg peak positions with the phases written in the legend.
}
\label{Reduced_TGA_XRD}
\end{figure}

In Fig. \ref{Reduced_TGA_XRD} we show the Ar-5\%H$_2$ reduction of the x = 0.25 variant in a TGA experiment performed at 450 $\degree$C revealing a mass loss of 5 wt\% corresponding to an oxygen loss of 2 O. Resulting in a reduction towards Sr$_{3}$Ni$_{1.75}$Al$_{0.25}$O$_{5}$ (Ni$^{+1.75}$), which is well captured by the XRD shown in Fig. \ref{Reduced_TGA_XRD} c) corresponding to the formation of the orthorhombic $Immm$ Sr$_{3}$Co$_{2}$O$_{5}$ type structure shown in Fig. \ref{Reduced_TGA_XRD} b) with the results summarized in Tab. \ref{tab:reduced_atom_pos}. The amount of decomposed parts found in powder XRD likely stems from a real reduction towards the more stable Ni$^{2+}$ in the crystals balanced by partial decomposition. Nonetheless our EDX results on these reduced crystals yield a similar oxygen content with Sr$_{2.98(1)}$Ni$_{1.81(2)}$Al$_{0.25(1)}$O$_{4.9(3)}$, which however has to be taken with a way larger errorbar as this result stems only from EDX. Notably, with the usual separation of domains typically found in the reduction of nickelates \cite{Puphal2021,Puphal2023} we find twinned single crystals visible in the single crystal XRD maps shown on the inset of Fig. \ref{Reduced_TGA_XRD} a).
The structural change upon reduction is clearly seen in the lattice constants (see Tab. \ref{tab:reduced_atom_pos}), where the $c$- and $a$-axis expand by 0.2 \AA, while the $b$-axis shrinks by the same value indicating the removal of in-plane oxygen along $b$. This is in line with other classical $T^{4+}$ RP, different to La$_3$Ni$_2$O$_7$ where the  $c$-axis shrinks by 1 \AA, while $a$, $b$ expand by 0.2 \AA \, \cite{Liu2022growth}.

\begin{table}[h]
\caption{Refined atomic coordinates of Ar/H$_2$ treated  Sr$_3$Ni$_{1.75}$Al$_{0.25}$O$_{7-\delta}$ found to crystallize in the orthorombic $Immm$ (\#71) structure, extracted from powder XRD ($a=3.603(2)$ {\AA}, $b=3.839(3)$ {\AA}, $c=20.442(14)$ {\AA}; Rwp$=7.00$, GOF$=0.95$)}
\label{tab:reduced_atom_pos}
\begin{tabular}{cccccc}
\hline
\textbf{Atom} & \textbf{Label} & \textbf{x} & \textbf{y} & \textbf{z} & \textbf{Occ} \\ \hline
Sr            & Sr1            & 1/2    & 1    & 0.1780(4)   & 1            \\
Sr            & Sr2            & 0    & 1/2    & 1/2   & 
1            \\
Ni            & Ni1            & 1/2    & 1    & 0.3996(8)    & 7/8        \\
Al            & Al1            & 1/2    & 1    & 0.3996(8)    & 1/8        \\
O             & O1             & 1/2    & 1    & 0.3150(3)    & 0.5415(7)       \\
O             & O2             & 1/2    & 1    & 1/2   & 
1            \\
O             & O3             & 1/2    & 1/2    & 3677(14)    & 
1            \\
\end{tabular}
\end{table}

\begin{figure}[h]
\centering
\includegraphics[width=1.0\columnwidth]{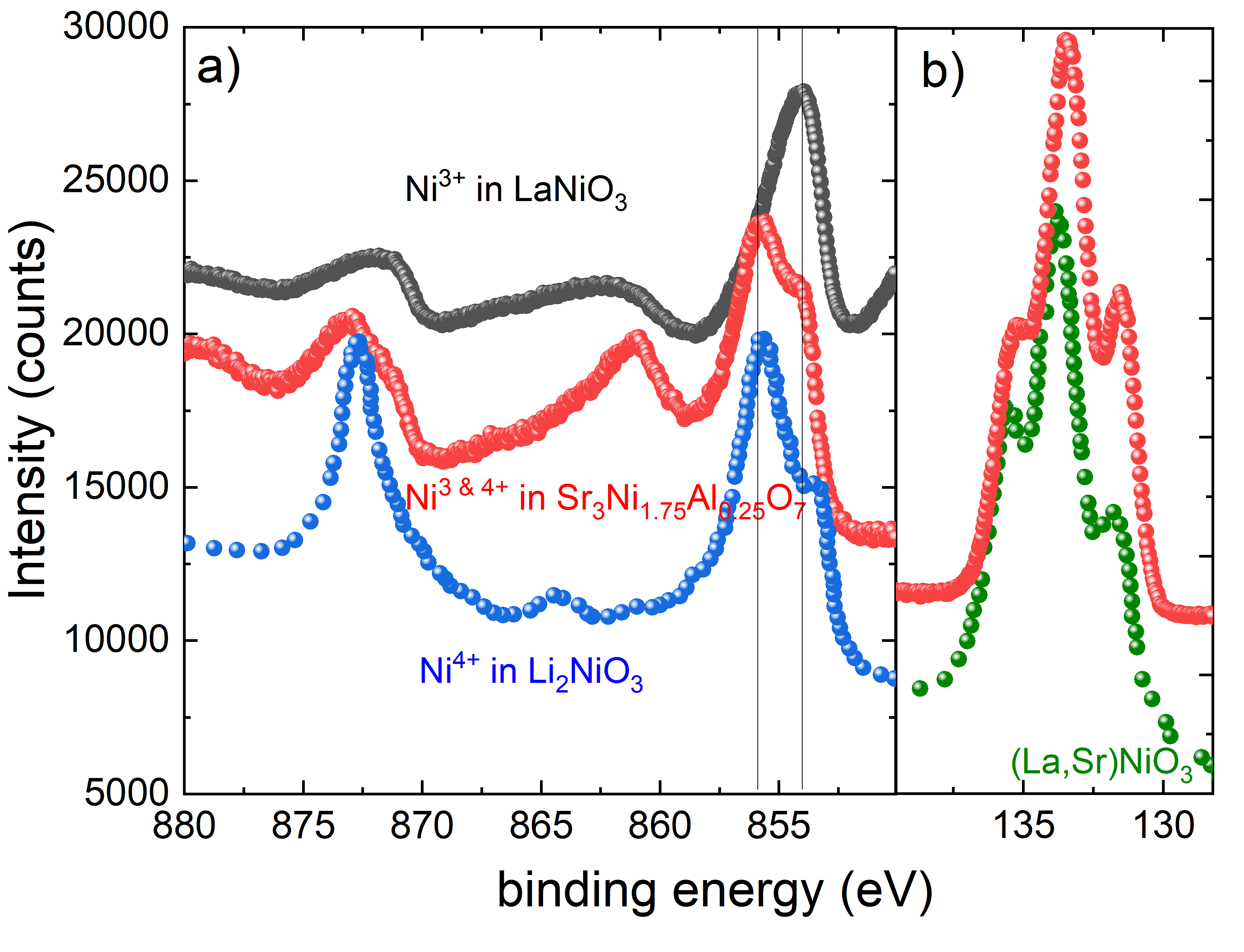}
\caption{\textbf{X-ray Photoemission Spectroscopy (XPS)} of Ni 2p and Sr 3d in Sr$_{3}$Ni$_{1.75}$Al$_{0.25}$O$_{7}$ compared to a XAS spectra of Li$_2$NiO$_3$ \cite{Bianchini2020} and our XPS of LaNiO$_3$, as well as Sr doped LaNiO$_3$ \cite{Puphal2023APL}.}
\label{XPS}
\end{figure}

For optimally grown Sr$_3$Ni$_{1.75}$Al$_{0.25}$O$_7$ we have in addition performed XPS measurements to confirm the unusual Ni$^{4+}$ state. In Fig. \ref{XPS} we show the Ni 2p and Sr 3d spectra, and compare it to literature results of comparable Ni$^{3+}$ and Ni$^{4+}$, which reveal that for our crystals we find a mixture of the two oxidation states. The strongest peak is at 855.6 eV originating from Ni$^{4+}$, with a shoulder at 854 eV. This shoulder is more intense compared to the one in Li$_2$NiO$_3$. Since it is at a similar binding energy to the main peak in LaNiO$_3$ were Ni is in 3+ oxidation state we conclude that there is a small amount of Ni$^3+$ in our Sr$_3$Ni$_{1.75}$Al$_{0.25}$O$_7$.
Sr 3d shows two doublets one with the Sr 3d$_{5/2}$ binding energy at 131.5 eV and a second one at 133.5 eV. The latter can be related to the instability of Sr on the surface probably reacting to Sr-oxide or hydroxide \cite{Wang2018X}.

\begin{figure}[tb]
\includegraphics[width=1.0\columnwidth]{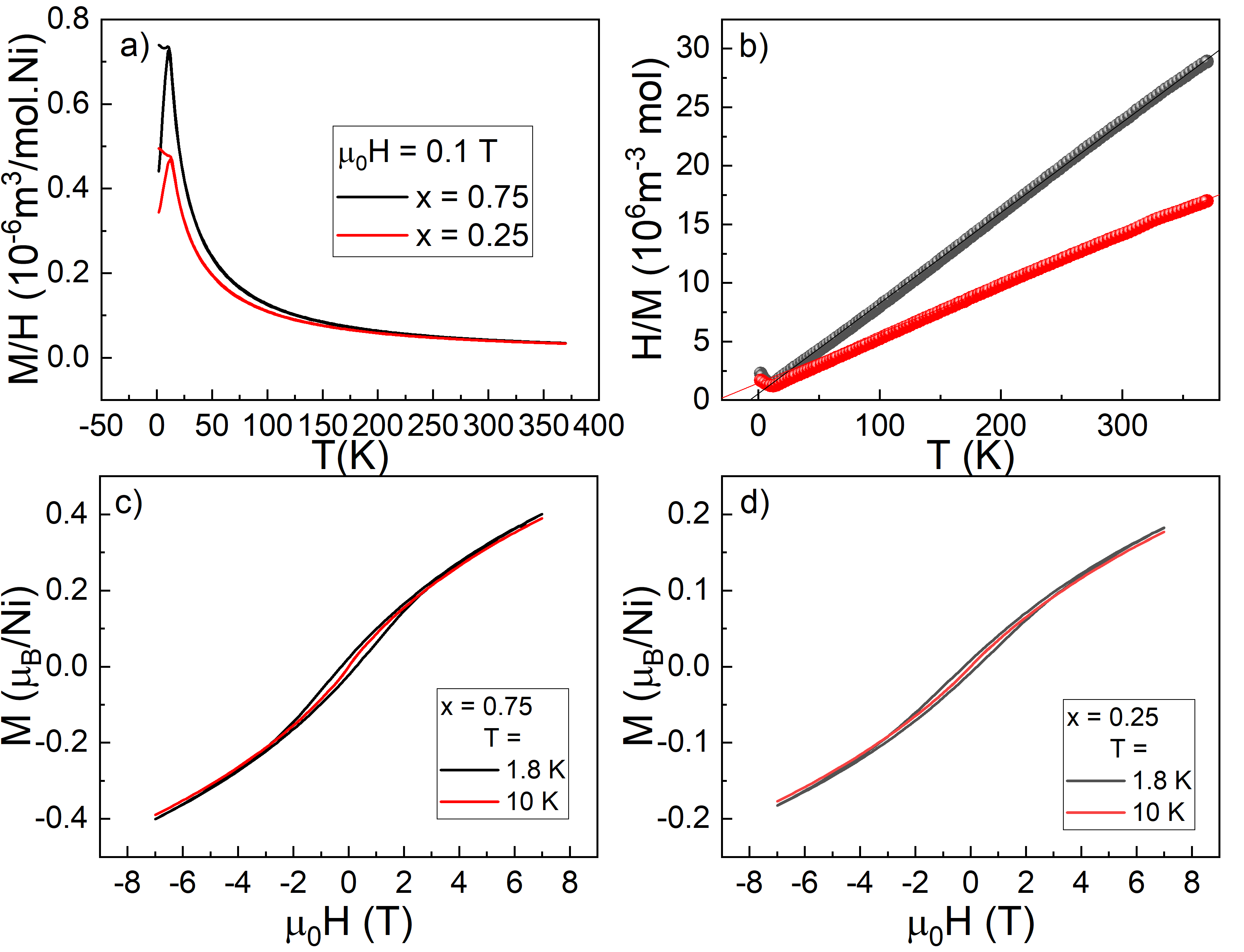}
\caption{\textbf{Doping dependence of the magnetism.} a) Magnetization versus temperature of a Sr$_3$Ni$_{1.25}$Al$_{0.75}$O$_7$ (x=0.75) and a Sr$_3$Ni$_{1.75}$Al$_{0.25}$O$_7$ (x=0.25) single crystals showing a magnetic transition at 11.5 K and 12.5 K. b) Inverse of the magnetization versus temperature with a Curie-Weiß fit. Magnetization versus field of c) Sr$_3$Ni$_{1.25}$Al$_{0.75}$O$_7$ (x=0.75) and d)Sr$_3$Ni$_{1.75}$Al$_{0.25}$O$_7$ (x=0.25)  at 1.8 and 10 K. 
}
\label{SQUID}
\end{figure}

\subsection*{Magnetic and electrical properties}
Finally, we investigate the physical properties of this classical RP-type nickelate phase. First we consider the spin state of Ni$^{4+}$. The low spin configuration has t$^0_{2g}$e$^6_g$, $S=0$ resulting in a nonmagnetic ion, while the high spin state would  have t$^2_{2g}$e$^4_g$, $S=2$. 

As discussed in the XRD section the inner Ni-O distance is extended to 1.984(4) \AA, while the others lie around the expected 1.9 \AA. The elongation of the inner oxygen bond hints at an energy lowering of z-axis-based energy levels and a high spin state. Notably as discussed in Ref. \cite{Abragam1986}, a dynamic Jahn-Teller effect is provided by the spectra of impurities belonging to the iron group, which occupy interstitial positions. Here, g-factors for 3d$^{6}$ vary from 3 to 3.4 resulting in a reduction of the effective moment. Due to the site-mixed nature of our nickelates, this scenario might play a role.

Our magnetic susceptibility measurements are displayed in Fig. \ref{SQUID}. Over all Al dopings the crystals reveal a magnetic transition at around 12 K, which subtly shifts with x. Curie-Weiß fits show a $\theta_W=-8$ and -50 K with a $\mu_{eff}=2.9\mu_\mathrm{B}$ and $3.87\mu_\mathrm{B}$ for x = 0.75 and x = 0.25 closer to the expected moment of Ni$^{3+}$ with $\mu_{eff}=3.9\mu_\mathrm{B}$, while Ni$^{4+}$ high spin has an effective moment of $\mu_{eff}=4.9\mu_\mathrm{B}$ (note, however, the possibility of a reduced g-factor). The Magnetization versus field of both x = 0.75 and x = 0.25 show a hysteresis at 1.8 K with a coercive field around 0.25 T, a remnance of $0.011\mu_B$, but even at 7 T the magnetization is far from saturation, indicating stronger exchange interactions than given from the Curie-Weiß fit. 

\begin{figure}[tb]
\includegraphics[width=1.0\columnwidth]{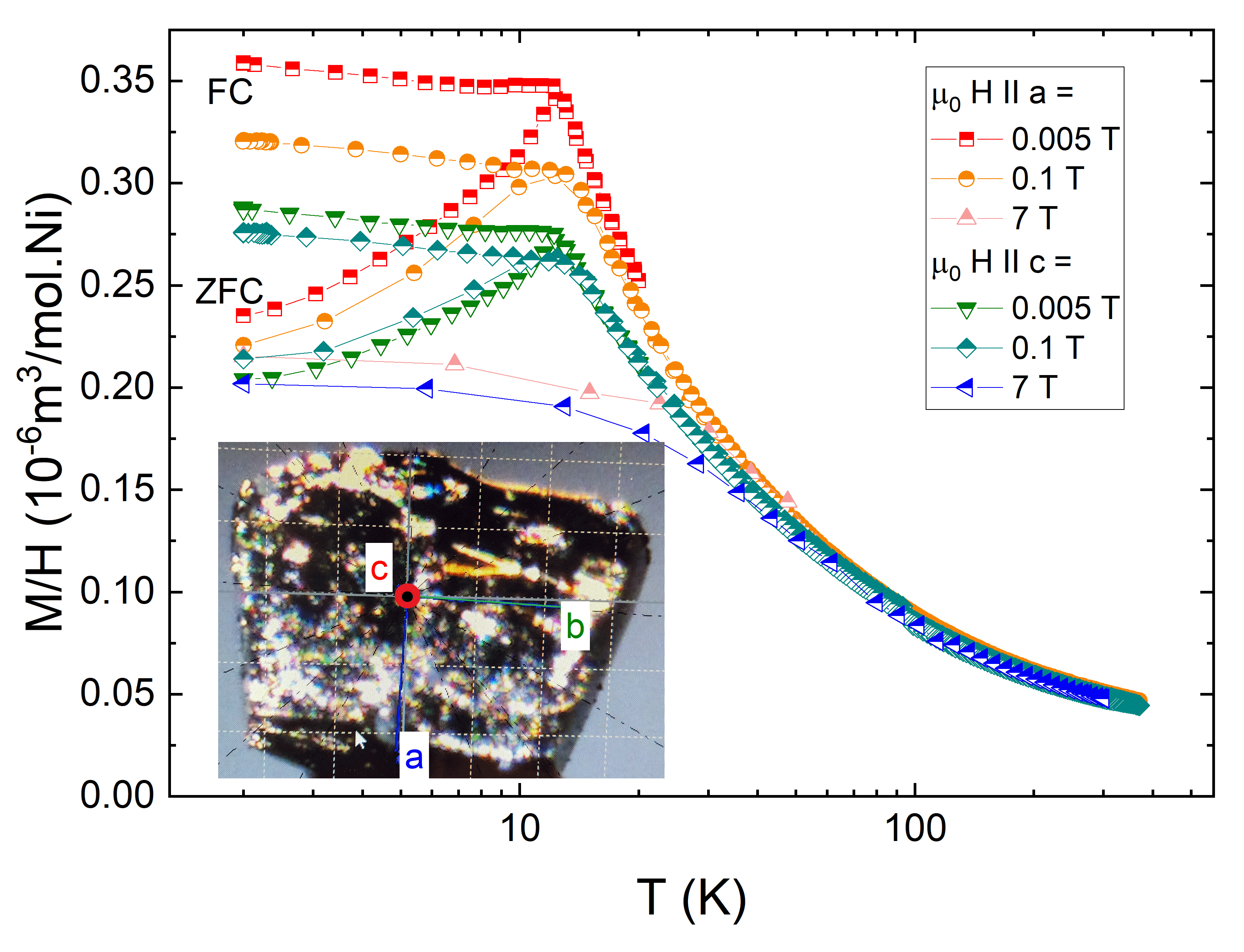}
\caption{\textbf{Anisotropy} and field dependence of the Magnetization versus temperature on a logarithmic scale to focus on the low temperature transition of a Sr$_3$Ni$_{1.75}$Al$_{0.25}$O$_7$ (x=0.25) single crystal in zero-field cooled (ZFC) and field cooled (FC) manner in fields of 0.005, 0.1 and 7 T. 
}
\label{Aniso}
\end{figure}

To investigate the field dependence and magnetic anisotropy we carefully checked a crystal by single crystal XRD and oriented it in this process. In Fig. \ref{Aniso} we see no major anisotropy concluding a spin-glass transition that is slowly suppressed by external fields and obviously originates from nonmagnetic Al site vacancies.

\begin{figure}[tb]
\includegraphics[width=1.0\columnwidth]{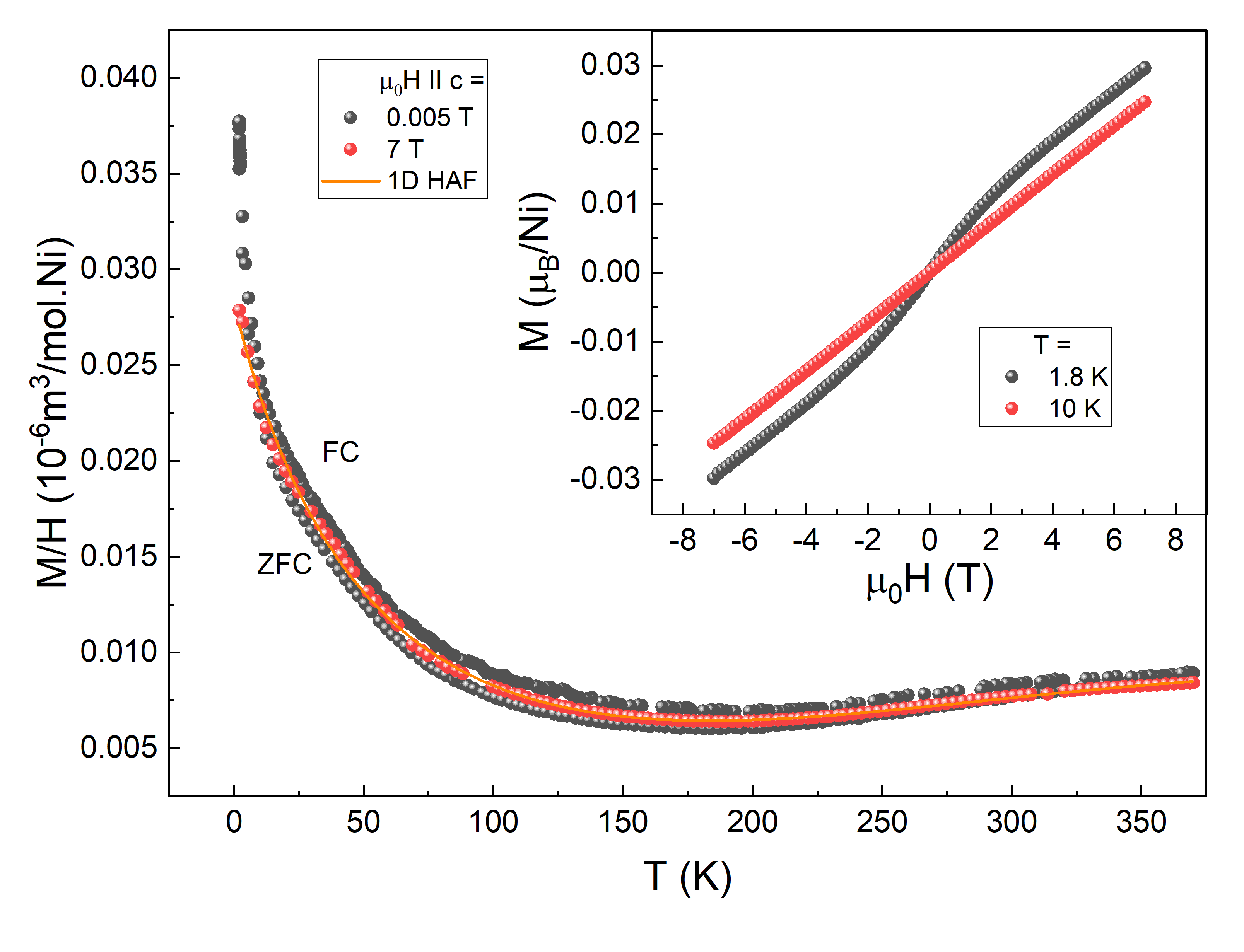}
\caption{\textbf{Sr$_9$Ni$_{7}$O$_{21}$} Magnetization versus temperature of a Sr$_9$Ni$_{7}$O$_{21}$ single crystals in a field of 0.005 T and 7 T. The inset shows the magnetization versus field at 1.8 and 10 K.
}
\label{SQUIDSNO}
\end{figure}

As the magnetic properties of the trigonal phase Sr$_9$Ni$_{7}$O$_{21}$ are yet unexplored, we performed a magnetization measurement of our rather inhomogenous crystals grown from off-stoichiometry, which however only contains nonmagnetic impurities. Notably from the stoichiometry we expect Ni close to 3+, likely in the high spin state t$^5_{2g}$e$^2_g$, $S=3/2$ and $J=9/2$. Looking at the structure shown in Fig. \ref{phases}, we find NiO face sharing chains, likely realizing a low dimensional quantum magnet indicated by a maximum seen in the susceptibility in Fig. \ref{SQUIDSNO}. Commonly found for Ni$^{3+}$ an effective $S=1/2$ state is realized (e.g. $R$NiO$_3$) this would allow us to describe our susceptibility versus temperature with a one dimensional Heisenberg Antiferromagnetic Chain (1D HAF) in addition with a curie term. 
Such a fit would result in an exchange coupling of 803(30) K. This likely overestimates J, which however certainly lies above our highest accessed temperature, i.e. 370 K. We observe a subtle zero-field cooled bifurcation indicating glassiness from structural disorder, and derivatives of the susceptibility show no transitions down to 2 K. The magnetization versus field at 10 K is linear, whereas the lowest temperature one shows a subtle kink around 1.4 K, likely from saturation of the paramagnetic contribution. Overall, the moment is small and far from saturation.

\begin{figure}[tb]
\includegraphics[width=1.0\columnwidth]{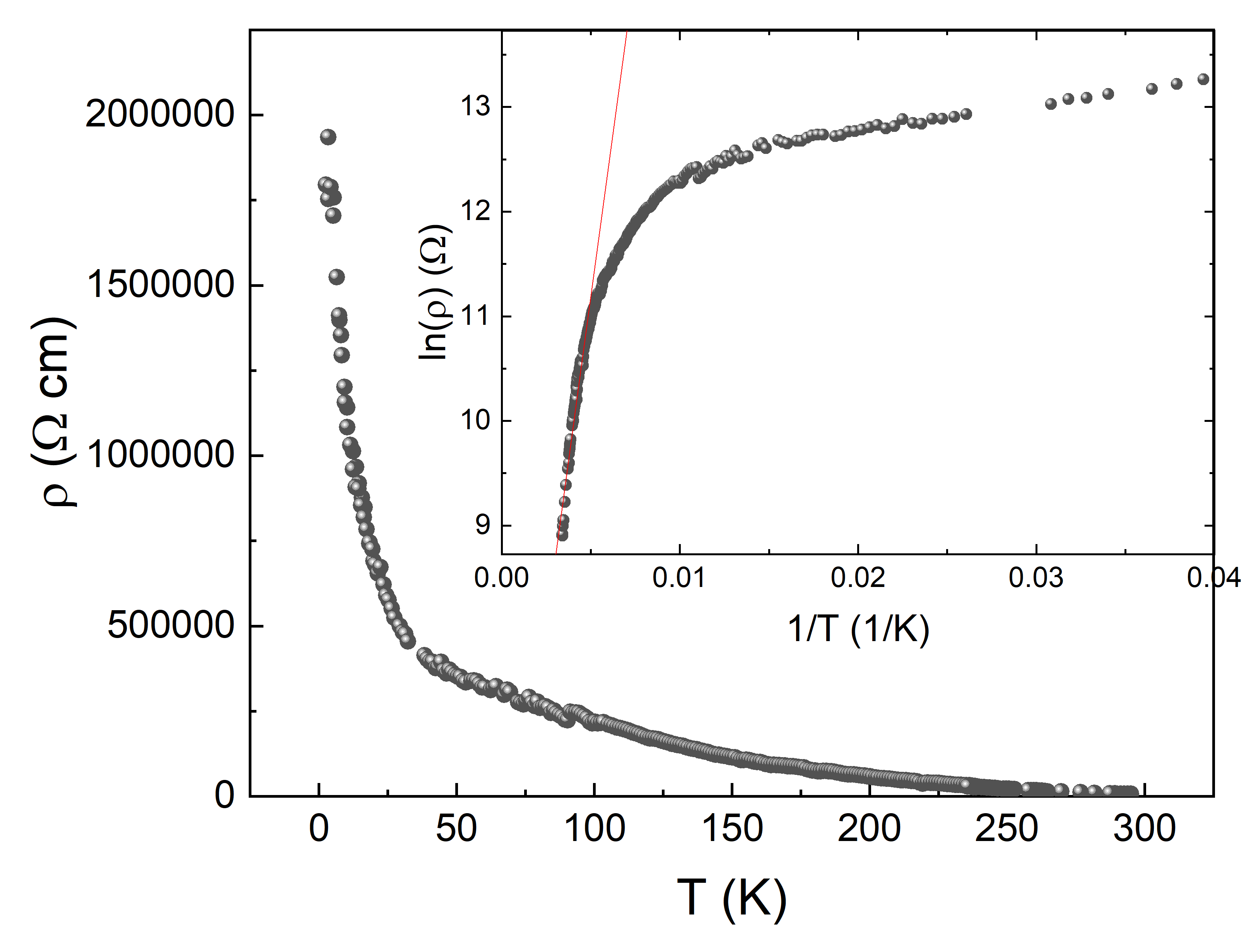}
\caption{\textbf{Electrical transport} versus temperature of Sr$_3$Ni$_{1.75}$Al$_{0.25}$O$_7$ showing highly insulating behaviour. The inset shows an Arrhenius plot.
}
\label{res}
\end{figure}

Sr$_9$Ni$_{7}$O$_{21}$ forms black crystals that are insulating. A quick four-point resitivity measurement failed as the resistance is above G\ohm.
In Fig. \ref{res} we show the resistivity of a x = 0.25 crystal, which reveals insulating behaviour likely induced by the Al doping. An estimation of the band gap energy via an Arrhenius law is applied at high temperatures. Here, all carriers are thermally excited across the band gap, and hence the electron and hole concentrations are equal. The fit yields $E_g=1250$ K using $\rho=\rho_0$exp$(E_g/(kT))$. Notably, all crystals are black, so future doping, or Al alternatives, might allow for the establishment of a metallic bilayer RP-type phase.

\section*{Discussion}
In Summary, we have presented the stabilization of bilayer n=2 Ruddlesden-Poppers type Nickelate Sr$_3$Ni$_{2}$O$_7$ via Al substitution on the Ni site, with the classical ($T^{4+}$) RP-type $n=2$ tetragonal $I4/mmm$ structure. With Sr$_3$Ni$_{1.75}$Al$_{0.25}$O$_7$ we find the optimal stoichiometry in terms of phase purity and crystallinity, where unlike La$_3$Ni$_{2}$O$_7$ no polymorphism and a wide phase stability range are observed. The oxygen content is determined to exactly balance the Al$^{3+}$ content with the formation of stable Ni$^{4+}$ using two independent techniques that lead to the final composition Sr$_3$Ni$_{1.75}$Al$_{0.25}$O$_{6.88}$. In particular, this is the fully oxidized sample, and no orthorhombic transition occurs even with annealing at 600 bar, in contrast to La$_3$Ni$_2$O$_7$. After reduction the system becomes orthorhombic $Immm$ Sr$_3$Ni$_{1.75}$Al$_{0.25}$O$_{5.125}$ with stable Ni$^{2+}$. Notably, the stabilization of Ni$^{4+}$ via Al substitution could help in Li-Ni-based battery applications and shows an enormous potential for stabilization of metastable phases. Our transport measurements show that the black crystals are insulating. We encourage theoretical calculations on the unique stabilization via Al$^{3+}$, while a try of Ti$^{4+}$ (not shown) instead resulted in the formation of a $n=1$ RP-type phase, and Al free synthesis gave crystals of the trigonal Sr$_9$Ni$_{7}$O$_{21}$ realizing face sharing one dimensional chains.  Hence, theoretical predictions of an optimal B site cation are desired. Although the crystals are insulating pressurization might enhance metallicity in these black crystals, as was shown for the insulating La$_3$Ni$_{2}$O$_6$ with the same crystal structure \cite{Liu2022growth}. Moreover, chemical doping will be pursued in the near future giving hope to novel nickelate superconductors.

\section*{Methods}
\subsection*{Synthesis}
For the synthesis of Sr$_3$Ni$_{1.75}$Al$_{0.25}$O$_7$ single crystals we utilize the optical floating zone (OFZ) technique under 9 bar oxygen partial pressure. First, the feed rods were prepared by conventional Solid-State methods using stoichiometric amounts of SrCO$_{3}$ (Sigma Aldrich, 99.995\%), NiO (Alfa Aesar, 99.998\%) and Al$_{2}$O$_{3}$ (Alfa Aesar, 99.995\%). The precursor powders were mixed using a ball mill and sintered in an aluminum oxide crucible at 1200 \textdegree C for 12 hours in air. The as-sintered powders were pressed into cylindrical rods (\(\phi\) = 4 mm, l = 100 mm) by an isostatic press (70 Mpa) using rubber forms and subsequently annealed at 1200 \textdegree C 12 hours in air. The rods were then installed in an Optical Float Zone Furnace (FZ-T-10000-H-III-VPR). Finally, single crystal growth was conducted equipped with 4x1000 W- halogen lamps. During the growth process, feed and seed rods were counter-rotated (24 rpm) in order to minimize the diffusion zone close to the crystallization surface. Oxygen gas flow (100 cc/min) was applied and an auxiliary pressure of 9 bar was introduced simultaneously for forming the oxidized phase. To obtain relatively large single crystals, the growth rate was stabilized to 4 mm/h with feeding at the same rate. Post-growth, the boule was cracked using a mortar and single crystals were selected for further analysis. Selected crystals have been annealed using a high pressure gas autoclave with a gold membrane compressor at 600 bar oxygen pressure and 600$\degree$C for 4 days.

\subsection*{Diffraction}
The crystallinity and phase purity of the as-grown single crystals were checked using XRD on powdered samples. Powder XRD measurements were performed on ground crystals in order to examine the presence of impurity as well as to determine the crystal structure and lattice parameters. XRD patterns were recorded at room temperature using a Rigaku Miniflex diffractometer with Bragg-Brentano geometry, Cu K${_\alpha}$ radiation and a Ni filter. Rietveld refinements were conducted with the TOPAS V6 software. Single-crystal diffraction was performed at room temperature using a Rigaku XtaLAB mini II instrument with Mo K${_\alpha}$ radiation. The data was analyzed with CrysAlis(Pro) and the final refinement was done using Olex2 with SHELX. 

\subsection*{Chemical composition}
XPS data were collected using a commercial Kratos AXIS Ultra spectrometer and a monochromatized Al K$_\alpha$ source (photon energy, 1486.6 eV). The spectra were collected using an analyzer with a pass energy of 20 eV. Due to the insulating nature of the sample a charge neutralizer was used and the C 1s was set to 284.8 eV (adventitious carbon) for binding energy calibration \cite{Biesinger2022}. XPS spectra were analyzed using the CASAXPS software. All samples were cleaved in a glove box, mounted via silver paint on a pin, and
transported under an inert atmosphere to the XPS chamber.
Energy-dispersive x- ray spectra (EDX) were recorded with a NORAN System 7 (NSS212E) detector in a Tescan Vega (TS-5130MM) Scanning electron microscope (SEM). Differential Thermal Analysis studies (DTA) and Thermo gravimetry (TG) was done using a Netzsch DTA/TG.
Via an Eltra ONH-2000 analyzer we determine the precise oxygen content, where the samples were placed in a Ni crucible, clipped and heated. A  carrier  gas  takes  the  oxygen, after which the oxygen reacts with carbon and  CO$_{2}$  is  detected  in  a  infrared-cell. Each measurement is repeated three times and compared with a standard. The quoted error bars give the statistical error.

\subsection*{Magnetic and electric properties}
Magnetic susceptibility measurements were performed using a vibrating sample magnetometer (MPMS 3, Quantum Design). Electrical transport measurements were carried out with a Physical Property Measurement System (PPMS, Quantum Design).

\section*{Author contributions}
P.P. conceived the project supervised the experiments and measured transport and magnetization. H.Y. grew the crystals, performed XRD measurements under the supervision of P.P.. O.C. and M.I. were responsible for the project management. XPS samples were prepared by P.P., measured and analyzed by K.K. and U.S.. H.Y. and P.P. wrote the manuscript with comments from all authors.

\begin{acknowledgements}
We thank Samir Hammoud for carrying out gas extraction studies. Part of the magnetic measurements are based on a VSM-SQUID of the solid state spectroscopy department at Max Planck Institute for Solid State Research, Heisenbergstrasse 1, 70569 Stuttgart, Germany.
\end{acknowledgements}

\bibliography{Literature}
\end{document}